\documentclass[epj]{svjour}
\usepackage{graphics}

\def\epem{${\rm e}^+ {\rm e}^-$}
\newcommand{\be}{\begin{equation}}
\newcommand{\ee}{\end{equation}}                                                                               
\newcommand{\bea}{\begin{eqnarray}}
\newcommand{\eea}{\end{eqnarray}} 

\begin{document}
\title{Investigating the QCD phase diagram with hadron multiplicities 
at NICA}
\author{F. Becattini \inst{1} \and R. Stock \inst{2}}
\institute{Universita' di Firenze and INFN Sezione di Firenze, Italy \and
Goethe University, Frankfurt am Main, Germany}
\date{Received: date / Revised version: date}
\abstract{
We discuss the potential of the experimental programme at NICA to investigate
the QCD phase diagram and particularly the position of the critical line at
large baryon-chemical potential with accurate measurements of particle 
multiplicities. We briefly review the present status and we outline the tasks 
to be accomplished both theoretically and the experimentally to make hadronic
abundances a sensitive probe.
\PACS{
      {PACS-key}{discribing text of that key}   \and
      {PACS-key}{discribing text of that key}
     } % end of PACS codes
} %end of abstract
\maketitle
%*****************************
\section{Introduction}
\label{intro}
%*****************************

The determination of the critical line of QCD is one of the principal goals of 
relativistic nuclear collisions. It is especially so in the energy range covered 
by the NICA experimental programme where one cannot rely on first principle
calculations and one would like to identify the supposed critical point. 
Indeed, the nucleon-nucleon centre-of-mass energy range $\sqrt s_{NN}$ between few 
and 10 GeV, corresponds, according to most extrapolations \cite{beca06,cleymans,andronic} 
to a baryon chemical potential at chemical freeze-out between 300 and 800 MeV, a 
range which is hardly accessible to lattice QCD owing the notorious sign problem. 

The two main questions to cope with, in the NICA energy range, are: 
\begin{enumerate}
\item{} has the colliding system thermalized and crossed the critical line?
\item{} what is the hadronic observable best suited to probe the crossing?
\end{enumerate}
In order to answer these questions, one can, of course, take advantage of the accumulated
evidence and knowledge about hadron formation in nuclear collisions at higher energy
as well as in elementary collisions. What we have learned from all the previous 
experience is that hadron formation is basically a universal local statistical process 
with some remarkable difference in the strangeness sector, whose phase space appears 
to be only partially filled in elementary collisions \cite{stock,becareview} \footnote
{It is worth pointing out here that the strangeness undersaturation is 
still observed in nuclear collisions at high energy but it can be accounted for 
by residual nucleon-nucleon collisions nearby the outer edge of the nuclear overlapping 
region (core-corona model \cite{werner,becacore}).}. 
The salient feature of relativistic nuclear collisions is thus the production - at 
hadronization 
- of a fully equilibrated hadronic system whilst in e.g. \epem or pp collisions 
strange particles do not saturate the available phase space. The deep reason of these 
behaviour is still unsettled and its meaning under debate, but for practical purposes 
this evidence may be used to probe the crossing of the critical line under the 
reasonable assumption that hadronization coincides with it. In other words, if the 
hadron production process in a nuclear collision was fully consistent with a picture 
of subsequent and independent elementary hadronic reactions (like in the hadronic 
transport models), hence with a - most likely - relatively large undersaturation 
of strangeness, there would be no reason to claim that the critical line has been 
overcome. 
%+++++++++++++++++++++++++++++++
If this was the case from some energy downwards, it would mean in fact that in that 
region, where a hadronic kinetic model is perfectly able to reproduce the data, 
critical line could not be located by studying hadron production in nuclear collisions. 
%++++++++++++++++++++++++++++++++

The hadronic observables which can be used to investigate the phase transition are
in principle many. It has been proposed to use fluctuations of conserved 
charges \cite{karschgroup,ratti} for they can be directly calculated in lattice QCD,
unlike hadronic multiplicities, hence comparable with measurements and used to
determine $T$ and $\mu_B$ if the system has reached a point of equilibrium. However, 
multiplicities are first moments and, as such, far more robust observables against 
spurious effects. Indeed, fluctuations are strongly affected by both the effect of
conservation laws \cite{goren} and finite acceptance \cite{koch}, so that the 
interpretation of measurements requires much care and the subtraction of non-
thermodynamic contribution can be a challenging task. For these reasons, we will 
focus on hadronic multiplicities and try to advocate them as still one of the best 
probes at our disposal to study and investigate the physical problem. We will 
summarize recent advances and outline what are, in our view, the next steps to be 
taken in the analysis to give a definite answer to the question. 

%***********************************************************
\section{Hadronization, statistical model and recent advances}
%***********************************************************

As has been mentioned, a statistical {\it ansatz} is able to satisfactorily reproduce 
the measured hadronic yields, both in elementary \cite{becaelem} and in relativistic 
nucleus-nucleus collisions \cite{various}. This has led to the formulation of the 
statistical hadronization model which, in a nutshell, assumes that hadrons are emitted 
from the fireball source at (almost) full chemical equilibrium. The reason of such a 
success, unexpected in elementary collisions, as well as the identity of the fitted 
temperature in all kinds of collisions, has been debated for a long time 
(see refs.~\cite{satz,becareview} for a summary). In practice, one can take advantage 
of this phenomenon to locate the parton-hadron coexistence line of QCD matter in the 
$(T,\mu_B)$ plane.
%--------------------------------------------------------------------
\begin{figure}
\resizebox{0.5\textwidth}{!}{\includegraphics{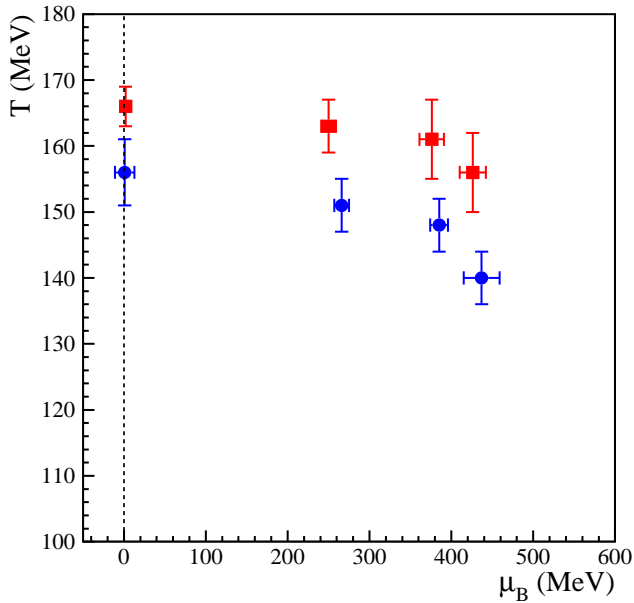}}
\caption{(color online). Temperature and baryon-chemical potential of the hadronic system
produced in central heavy ion collisions. Blue dots indicate chemical freeze-out 
points whereas red squares show the reconstructed LCEP's using UrQMD afterburner 
(taken from table in ref.~\cite{freez2}) }.
\label{fig1}       
\end{figure}
%--------------------------------------------------------------------

The temperature determined by fitting the hadronic multiplicities is actually the 
one at which hadrons cease inelastic interaction, the so-called 
"chemical freeze-out" temperature. In principle this may differ from the QCD transition 
temperature if hadrons, after their formation, keep interacting inelastically. This 
is, clearly, not the case in elementary \epem annihilation to hadrons but it could 
become relevant in the high multiplicity final state of AA collisions. Different 
reactions could then freeze-out at different times, in inverse order of inelastic 
cross section, so that this stage of the fireball source expansion, dubbed as "afterburning", 
would generally imply deviations from full chemical equilibrium of the hadronic 
species \cite{bass}. In the standard statistical model analysis such effects were 
assumed to be negligibly small, and that, therefore, the temperature and baryon 
chemical potential yielded an ideal snapshot of the fireball dynamical trajectory, 
at or near QCD hadronization.

However, recently, we have demonstrated that afterburning, albeit being a correction 
to the leading statistical behaviour, ought to be taken into account as it implies
a considerable improvement in the fit quality and - more importantly - a sizeable 
increase of the chemical freeze-out temperatures \cite{freez1,freez2,freez3}. 
Similar improvements in the agreement between data and model were reported in
ref.~\cite{hybrid}. The underlying physical picture is as follows: the hadronization 
process (likely to coincide with the crossing of critical line) entails a primordial
chemical equilibrium of multiplicities which is then distorted by the following stage 
of hadronic inelastic collisions before chemical freeze-out. This stage has a sizeable 
dependence on the geometry of the collision, the colliding nuclei and on their energy. 
The goal of the fit is to reconstruct the primordial chemical equilibrium point 
(defined as the {\em latest chemical equilibrium point} LCEP). The LCEP's are likely 
to coincide with the hadronization, hence with the a point on the critical line and 
if the fit is succesfull, with no evidence of an extra strangeness suppression other 
than that related to corona collisions, we can reasonably state that the critical 
line has been overcome. 

%+++++++++++++++++++++++++++++++++++++++++++++++++++++
The core method of the aforementioned analyses - described in detail in 
refs.~\cite{freez2,freez3} was to calculate particle multiplicities at the LCEP with 
the Cooper-Frye prescription and, thereafter, to let particle interact till freeze-out 
with the UrQMD transport code \cite{urqmd}, to get a quantitative estimate of the 
influence of the rescattering stage.
%+++++++++++++++++++++++++++++++++++++++++++++++++++++
This approach worked very well for the most central collisions from LHC down to SPS 
energy (where data collected by the NA49 experiment were used \cite{na49}), providing 
an indication that the critical line has indeed been crossed down to $\sqrt s_{NN} 
= 7.6$ GeV, although a detailed analysis of the corona collisions is still missing. 
In fig.~\ref{fig1}) we show our reconstructed LCEP's along with the uncorrected chemical 
freeze-out points. The LCEP's points define a sizeably flatter curve in the QCD phase 
diagram whose curvature seems to be in a better agreement with lattice calculations 
\cite{pisa,wuppertal,hotqcd,bari}.

One can wonder whether it could be extended further down, where at 
some point it will have to fail, signalling that the critical line has not been attained.
Unfortunately, the data from AGS in this respect does not allow firm conclusions, 
because the measured multiplicities are few and the full phase space measurements 
are even fewer. The present statistical model fits in this region \cite{cleymans}
in the authors view, are thus not as reliable as those at energies higher than 
AGS, that is $\sqrt s_{NN} > 5$ GeV. For this reason, a detailed exploration of the energy 
range 2-10 GeV is compelling and accurate measurements of hadronic multiplicities 
are needed.

%************************************************
\section{A task list}
%************************************************

In order to establish an interval of energies where transition line has possibly 
been crossed, it is necessary to have a set of very accurate measurements on the 
experimental side and very accurate calculations on the theoretical side. Unfortunately, 
on both sides, it is not always possible to achieve the desired accuracy, nevertheless 
it is worth listing here the problems that will be encountered and the future demands.

From the experimental viewpoint, a measurement of the multiplicities of many species 
in full phase space besides midrapidity will be necessary. It is known that a cut at 
midrapidity artificially enhances strange particles and that it introduces more 
uncertainties on the quantitative effects of canonical suppression. This is a known
problem in, e.g. the RHIC beam energy scan, where the use of exact strangeness 
conservation in the fit of midrapidity yields changes the slope of the chemical 
freeze-out temperature as a function of centrality \cite{bes}. Finally, for the 
sake of systematic consistency while changing the incident energy, it is essential 
to maintain the same set of hadronic species, covered in the analysis (ideally extending 
down to the small $\Omega$ hyperon yield).

In order to tune the hadronic models and to take into account corona effects, a 
measurement of multiplicities in pp (an possibly np and nn) collisions at the same 
energies and within the same kinematical windows of the heavy ion run will also 
be necessary. Of course, in this respect, the NA61/SHINE \cite{shine} pp data will 
be helpful, if taken at the same energy. 
%++++++++++++++++++++++++++++++++++
For the same purpose, a measurement over a sufficiently large centrality range 
in the nuclear collisions will be crucial to check that hadron production is fully
understood.  
%++++++++++++++++++++++++++++++++

From a theoretical viewpoint, the endeavour will be a major one. The hadronic transport
models will have to be coupled to the statistical hadronization (at high energy this
happens through the Cooper-Frye prescription) as well as be used in a standalone 
mode \cite{urqmd,epos,ampt,phsd}. Statistical hadronization normally uses the hadron-resonance 
gas implementation, which amounts to disregard the contribution of the non-resonant 
hadronic interaction as well as higher order corrections. This is expected to be 
(and confirmed by lattice QCD) a very good one at low $\mu_B$, but it is not yet clear 
if it will be as good at large $\mu_B$, where the continuum may play a major role. 
Going beyond the hadron-resonance gas in the statistical model will be possibly needed, 
what will require a major theoretical and computational effort.

%************************************************
\section{Conclusions}
%************************************************

The precise measurement of hadronic multiplicities entailed a substantial progress 
in understanding hadronization and its relation with QCD over the last twenty years
and it will continue to do so in the future. Therefore, we think that it will be a 
must for the NICA experimental programme to achieve complete and accurate sets of 
measurements of as many hadronic species as possible, at various closely spaced 
energies. We wish to note, in passing, that a relatively narrow sequence of LCEP's 
points in the $T-\mu_B$ plane, gathered at closely spaced successive incident 
energies, might also exhibit a non-trivial, non-monotonic behaviour, as a consequence 
of the expansion trajectory focussing effect due to a critical point of QCD occuring 
in the thus covered interval of $\mu_B$ This interesting proposal \cite{lacey} has 
not been further elaborated theoretically, let alone tested experimentally.

%**************************************************************

\end{document}